\documentclass{aa}
\usepackage{graphicx,natbib}
\usepackage{natbib}
\bibpunct{(}{)}{;}{a}{}{,} 
\usepackage{graphicx}
\usepackage{txfonts}
\newcommand{\p}{\partial}
\newcommand{\nab}{{\bf \nabla}}

\newcommand{\BB}{\mbox{\bf B}}
\newcommand{\vv}{\mbox{\bf v}}
\newcommand{\be}{\begin{equation}}
\newcommand{\ee}{\end{equation}}

\begin{document}

\title{Accretion funnels onto weakly magnetized young stars}

\author{Bessolaz, N. \inst{1} \and Zanni, C. \inst{1} \and Ferreira,
  J. \inst{1} \and Keppens, R. \inst{2,3,4} \and Bouvier, J. \inst{1}}

\institute{Laboratoire d'Astrophysique de Grenoble, Universit\'e
  Joseph Fourier, CNRS UMR5571, France \and 
           Centre for Plasma Astrophysics, K.U.Leuven, Belgium \and 
           FOM Institute for Plasma Physics, Rijnhuizen, the Netherlands \and 
           Astronomical Institute, Utrecht University}

\abstract{}
{
We re-examine the conditions required to steadily deviate an accretion
flow from a circumstellar disc into a magnetospheric funnel flow onto
a slow rotating young forming star.}
{
New analytical constraints on the formation of accretion funnels flows
due to the presence of a dipolar stellar magnetic field disrupting the
disc are derived. The Versatile Advection Code is used to confirm these constraints numerically. Axisymmetric MHD simulations are performed, where a stellar dipole field enters the resistive accretion disc, whose structure is self-consistently computed.}
{
The analytical criterion derived allows to predict a priori the
position of the truncation radius from a non perturbative accretion
disc model. 
Accretion funnels are found to be robust features which occur
below the co-rotation radius, where the stellar poloidal magnetic
pressure becomes both at equipartition with the disc thermal pressure
and is comparable to the disc poloidal ram pressure. 
We confirm the results of Romanova et al. 2002 and find accretion
funnels for stellar dipole fields as low as 140 G in the low
accretion rate limit of $10^{-9} \mathrm{M_\odot.yr^{-1}}$.
With our present numerical setup with no disc magnetic field, we found no evidence of winds, neither disc driven nor X-winds, and the star is only spun up by its interaction with the disc.}   
{
Weak dipole fields, similar in magnitude to those observed, lead to the development of accretion funnel flows in weakly accreting T Tauri stars. However, the higher accretion observed for most
T Tauri stars (${\dot M} \sim 10^{-8} \mathrm{M_\odot.yr^{-1}}$) requires either larger stellar field strength and/or different magnetic topologies to allow for magnetospheric accretion.}


\maketitle

\section{Introduction} 

Classical T-Tauri stars (CTTS) are magnetically active, show evidence for circumstellar accretion discs, and can have mean
photospheric magnetic field magnitudes around 1kG
\citep[e.g. ][]{jk}. Such a strong stellar field is enough to disrupt the inner accretion disc,
provided that one really measures the large scale magnetic field and
not only local strong multipolar components from starspots. However,  
this does not seem to be the case since recent polarimetric measurements
\citep{val} indicate a weak dipolar component lower than 200 G.
Moreover, observations show evidence for non-direct accretion.  
Inverse P-Cygni profiles with strong redshift absorption wings are
indicative of polar accretion near free-fall velocities along
magnetospheric field lines from the inner disc edge 
\citep{edw, Bou, Bou2}.

Although the magnetic field structure of these stars is probably
complex \citep{greg}, dynamical models of star-disc
interaction usually assume an aligned dipole
field, to simplify the analytical work. Under this assumption, stellar
field lines threading the Keplerian disc below the co-rotation
radius $r_\mathrm{co}=(GM_*/\Omega_*^2)^{1/3}$, where $\Omega_*$
is the stellar angular velocity, would lead to a spin-up of the star,
versus those beyond $r_\mathrm{co}$, to a spin-down. The radial extent of
angular momentum exchange between the star and the disc is then
determined by (1) the disc truncation radius $r_\mathrm{t}$, where the magnetic
dipole diverts the radially accreting flow to funnel flows; and (2) an
outer radius $r_\mathrm{out}$ beyond which no more stellar field lines are
connected to the disc. In this framework, a star-disc interaction
occurring on a large radial extension \citep[as proposed by][]{ghosh, cc,
  Armi96}, may lead to a disc-locking situation where the star
remains at a slow rotation rate,
despite accretion. On the other hand, it has been argued that this scenario is
unlikely, since the stellar field lines would be opened up by 
differential rotation until severing this causal link
\citep[for a recent discussion on that issue]{Aly90,lov95, matt}. 
The outcome of this latter scenario would be a star-disc interaction 
limited to a small radial extension around the disc truncation radius. 
Many theoretical models then assume that the disc inner edge should be
close to the co-rotation radius \citep{ko91, shu} for the sake of
angular momentum equilibrium.

For a decade, many numerical works have investigated the star-disc
interaction issue \citep{hayashi96, kuker,Long05}. 
However, the formation of accretion curtains seems
difficult to reproduce. Although \citet{Miller97} found such
polar accretion in the case of a kG dipolar stellar field 
associated with a disc field in the same direction to the latter, 
\citet{rom}  were the first to demonstrate magnetospheric accretion 
along stellar field lines for kG pure dipolar field in axisymmetric simulations
and next by performing 3D simulations \citep{rom03}.

In this paper, we address the issue of the disc truncation radius and
its localisation as a function of the disc (accretion rate) and stellar
(dipole field) parameters. In Sect.~2 we provide analytical
constraints for driving steady accretion funnels and derive an estimate of the
position of the truncation radius. We then use numerical MHD
simulations in Sect.~3 to verify this prediction. We confirm
magnetospheric accretion for a slowly rotating star with an inner
disc hole and a weak stellar magnetic field compatible with observations
of weak accretors.

\section{The disc truncation radius}

For a given accretion disc model, predicting where the truncation by the
stellar magnetosphere will occur is an important issue. 
Different estimates were given in the literature.
We consider here a pure dipolar field with a strength at the stellar surface in the equatorial
plane equal to $B_*$.

A first dimensional estimate expresses the truncation radius $r_\mathrm{t}$ 
in the form $r_\mathrm{t} = k r_\mathrm{A}$ where the Alfv\'en radius 
\be
r_\mathrm{A} = \left( \frac{B_*^4R_*^{12}}{2GM_*\dot{M_a}^2} \right)^{1/7}
\label{eq:alrad} 
\ee
is a characteristic length which can be derived by equating the ram pressure of a free-falling
spherical envelope with the magnetic pressure of a dipolar field \citep{Elsner77}.
Different estimates of the adimensional coefficient $k$ have been
given in the literature, ranging from
0.5 \citep{ghosh, ko91, Long05} to $\sim 1$ \citep{Arons93, Ostr95, Wang96}. 

A second criterion \citep[and references therein]{cc, Armi96, matt} states
that accretion funnels will take place when the magnetic torque due to
the stellar field matches the ``viscous" or turbulent torque, namely
\be
\mathrm{
- J_r B_z \simeq \frac{B_\phi^+ B_z}{\mu_o h} =  \frac{1}{r^2} \frac{\partial}{\partial r} \eta_v r^3 \frac{\partial \Omega}{\partial r} \simeq - \alpha_v \frac{P}{r }
}
\ee
 where $B_\mathrm{\phi}^+$ is the toroidal field at the disc surface,
 $h$ the disc scale height, $P$ the disc pressure and
 $\alpha_\mathrm{v}$ the \citet{ss} parameter. 
This criterion (hereafter A) gives an upper limit for the truncation radius, as it only defines a radius where the star-disc interaction starts to affect accretion. This maximal truncation radius $r_\mathrm{t,max}$ occurs when the plasma beta, defined as $\mathrm{\beta = 2\mu_oP/B_z^2}$, becomes $\mathrm{\beta= 2q/(\alpha_v
\varepsilon)}$ where $\mathrm{q= | B_\phi^+/B_z|}$ is a measure of the magnetic
shear and $\varepsilon=h/r$ is the disc aspect ratio. In a thin Keplerian
accretion disc, one gets $\beta \gg 1$ at $r_\mathrm{t,max}$ since q is
close to unity to avoid opening of the magnetosphere \citep[e.g.][]{matt}. 

A third criterion represents a more conservative approach, and gives only a lower limit $r_\mathrm{t,min}$. It states that accretion funnels take place when accretion is no longer possible because of the overwhelming field strength. 
It is usually written \citep{Kold02, rom}
\be
\mathrm{
\frac{B^2}{\mu_o} = \rho v^2 + P
\label{eq:rtmin}
}
\ee 
where $\mathrm{v}$ is the total speed. Note that this criterion
  is not predictive since we cannot calculate a priori the dominant
  azimuthal velocity $\mathrm{v_\phi}$ as it is itself an outcome of
  the star-disc interaction. Nevertheless, if one wishes to provide an
  estimate of the truncation radius, then we can use instead
  $\mathrm{B_z^2/\mu_o = \rho v_\phi^2}$, where $\mathrm{v_\phi = v_K=
    \sqrt{GM_*/r}}$ and a negligible thermal
  pressure (both approximations are valid in a thin disc). This will be
  our criterion B. At the truncation radius estimated as such, the
  plasma beta is $\mathrm{\beta = 2 \varepsilon^2 \ll 1}$.  

From the previous discussion, it appears quite obvious that $\beta
\sim 1$ is a better approximation for $r_\mathrm{t}$.
The stellar magnetic field, which is bound to become dominant in the magnetosphere, must first favour accretion, i.e. the magnetic torque must be negative. If this is not the case, namely if  $\mathrm{r >  r_{t,max} > r_{co}}$,  the disc material is radially expelled. This is the ``propeller"
regime as studied e.g. by \citet{ustyugova06} and references therein. 
Accretion thus implies $r_\mathrm{t} < r_\mathrm{co}$, with stellar
magnetic field lines as leading spirals. Below co-rotation, accretion
will proceed quite naturally thanks to both the viscous and the
stellar torques. There are then two more independent constraints that
must be fulfilled to produce steady funnel flows. 
First, the accretion flow must be prevented by the presence of the magnetosphere. The simplest way to express this is to require that the magnetic poloidal pressure balances the accretion ram pressure
$\mathrm{\rho u_r^2}$. This defines a radius $r_{bf}$ where 
\be
\mathrm{\beta \simeq m_s^{-2}}
\label{rbf}
\ee
where $m_s = u_r/C_s$ is the sonic Mach number measured at the disc midplane. 
Now, at radii  $\mathrm{r < r_{t,max}}$, accretion is mainly
due to the stellar torque and $\mathrm{m_s = 4q/\beta \gg \alpha
  \varepsilon}$: material is accreting to the star much faster than in
the outer accretion disc.  
Second, material at the disc midplane must be lifted and loaded onto
the stellar field lines. With a dipole field configuration, such a vertical motion can only be due to a vertical 
thermal plasma pressure gradient. It therefore requires that the magnetic field
compression is not too strong. This leads naturally to an equipartition, $\beta \sim 1$ \citep[as already proposed by][]{pr72, Aly80}.  

One important point to note is that once a large scale magnetic field is close to equipartition in an accretion disc, it is able to deviate a large fraction of the disc plasma from its radial motion to a vertical one. 
This has been shown with the calculations of magnetized accretion-ejection structures by \citet{fp95} and confirmed by numerical MHD simulations \citep{CK, zanni}. 
In one sense, making funnel flows involves the same physics as loading mass in magnetized jets. 
In fact, almost all of the disc mass can be lifted and loaded onto the field lines when $\beta \sim 1$, depending mostly on the field bending \citep[see Fig.~3 in][]{fp95}. 
This is why we assume, for finding a simple analytical criterion, that the disc truncation radius $r_t$ is close to the radius where $\beta \sim 1$, namely $r_t \sim r_{bf}$. 
Using the above estimates, we derive the following two constraints: 
\begin{equation}
\beta \sim 1 \ \ \ \     \mbox{and}    \ \ \ \   m_s \sim 1.
 \label{eq:cond} 
 \end{equation}
that must be fulfilled in order to provide steady state funnel
flows. For a dipole field, this translates into a theoretical truncation radius 
\begin{equation}
\frac{r_\mathrm{t,th}}{R_*} \simeq 2\ m_s^{2/7} B_*^{4/7} \dot M_a^{-2/7} M_*^{-1/7}
R_*^{5/7} 
\label{eq:rt} 
\end{equation}
where the stellar field $B_*$ has been normalised to 140 G, disc
accretion rate $\dot M_a$ to $10^{-8}$ M$_\odot$ yr$^{-1}$, stellar
mass to 0.8 M$_\odot$ and stellar radius to 2R$_\odot$. Note that these are all typical values for CTTS while the chosen value of the magnetic stellar field  
$B_*$ is consistent with observations of dipole fields in such objects \citep[see][and references therein]{val,Bou07}. 
It is clear that the conditions given by Eq. (\ref{eq:cond}) state that the accretion speed close to the base of the accretion funnel is of the order of the Alfv\'en speed. Indeed Eq. (\ref{eq:rt}) 
can be rewritten as $r_\mathrm{t,th} \sim m_s^{2/7} r_\mathrm{A}$, where the Alfv\'en radius for spherical accretion $r_\mathrm{A}$ was defined in Eq. (\ref{eq:alrad}).
On the other hand, the co-rotation radius is $r_{co}/R_* = 7.8\ M_*^{1/3}R_*^{-1}P_*^{2/3}$ for a typical 8-day stellar period $P_*$, which provides  
\begin{equation}
\frac{r_\mathrm{t,th}}{r_{co}} \simeq 0.25\ m_s^{2/7} B_*^{4/7} \dot M_a^{-2/7} M_*^{-10/21}
R_*^{12/7}  P_*^{-2/3}
\end{equation}
Remarkably, taking typical values for CTTS with a low magnetic dipole
field gives a theoretical truncation radius not only smaller than the
co-rotation radius but also consistent with observations of inner disc
holes \citep{naj}. This is in strong contrast with unobserved large
scale kG fields usually used in the literature. This implies that the 
formation of funnel flows should {\it always} spin up the star unless 
(i) enough stellar field lines remain connected to the disc beyond 
$r_{co}$ and/or (ii) ejection of stellar angular momentum is indeed 
taking place. In the following section, 2.5D numerical MHD
simulations are used to show that the conditions in (\ref{eq:cond})
are indeed those prevailing at the disc inner edge. If this is true, 
then we should recover a truncation radius located at our theoretical 
estimate (\ref{eq:rt}) for a low dipole field of 140~G.

\section{Numerical experiments}

\subsection{Equations and numerical setup}

We use the VAC \footnote{http://www.phys.uu.nl/$\sim$toth/} code
\citep{toth} to solve the full set of axisymmetric dimensionless 
resistive MHD equations (with the magnetic permeability
$\mu_0=1$) in cylindrical coordinates (r,z): 

\begin{equation}
\frac{\p \rho}{\p t} + \nab\cdot\left(\rho \vv_p\right) = 0,
\label{mhd1}
\end{equation}

\begin{equation}
\frac{\p \rho\vv}{\p t} + \nab\cdot\left(\rho \vv \vv -\BB\BB\right) +
\nab\left(\frac{B^2}{2}+ P\right) = - \rho\nab\Phi_G,
\label{mhd2}
\end{equation}

\be 
\frac{\p \BB}{\p t} + \nab\cdot\left(\vv\BB - \BB\vv\right) = \nab\times\left[\eta
\left(\nab\times \BB\right)\right],
\label{mhd3}
\ee

\be
\frac{\p E}{\p t} + \nab\cdot\left[\left(E+\frac{B^2}{2}+ P\right)\vv-\BB\left(\BB \cdot
\vv\right)\right] =
\eta J^2 - \BB \cdot \left(\nab \times \eta {\bf J}\right) ,
\ee
where $\rho$ is the plasma density, $\vv_p$ the
polo\"{\i}dal velocity, $\vv$ the total velocity, $\BB$ the magnetic field, $P$ the thermal
pressure, $\eta$ the magnetic resistivity, ${\bf J} = \nab\times\BB$
the current density, $\Phi_G=-\frac{GM_*}{(r^2+z^2)^{1/2}}$ the gravity
potential created by the central star and
$E=\frac{P}{\gamma-1}+\rho\frac{\vv^2}{2}+\frac{\BB^2}{2}+\rho \Phi_G$ the total 
energy density. The gravity is treated as a source term in the momentum
and energy equations of VAC. 

Time evolution is done with a conservative second order accurate total variation diminishing Lax Friedrichs scheme with minmod
limiters applied on primitive variables, except for density where  a
van Leer limiter is used instead to better resolve contact
discontinuities. Powell source terms are used to ensure the divergence
free property of the magnetic field and the code has been modified so
as to compute only the deviations from the dipolar component
\citep{Tanaka94, Powell99}. 
The splitting technique has three main advantages. First, it is crucial
to properly represent an initial force-free configuration numerically. Second,
since the total conserved energy contains only the energy associated
with the deviation from the dipolar field, this method improves the computation 
of the thermal energy in low $\beta$ regions. Third, when used in association
with the Powell method, it helps control the divergence of the magnetic field,
since only the deviation from the background field is used to calculate the divergence
and the Powell source terms. The divergence method used here gives 
$|\nabla\cdot\BB|\Delta R \le 0.01 |\BB|$ 
but simulations done with PLUTO \footnote{A description of this
code can be found in \citet{mignone}.}  and a constraint-transport scheme 
show that our results are not strongly modified by this method (Zanni et al, in prep). 

\subsection{Boundary conditions}

Boundary conditions take symmetric and asymmetric conditions for the
axis and the disc midplane, while continuous (outflow) conditions are
used at the outer edge. At the inner edge corresponding to the stellar
surface, a linear extrapolation of density and pressure is made. 
After using free extrapolation for the poloidal velocity, this latter 
is forced to be parallel to the total poloidal magnetic field. 
The poloidal field is fixed, contrary to \citet{rom}.
We use a boundary condition on the toroidal magnetic field designed by
Zanni et al. (in prep.). It allows us to derive $B_\phi$ by forcing the magnetic surfaces to locally rotate at the stellar velocity. 
In practice, the radial derivative of the toroidal field is computed from the angular momentum conservation equation where the temporal derivative giving the local acceleration is replaced by

\begin{equation}
\rho \frac{\partial}{\partial t}(\Omega r)= \rho \frac{(\Omega_{*} r+v_p B_\phi/B_p
-\Omega r)}{\Delta t}
\end{equation}
where the timescale $\Delta t = \Delta R / V_\mathrm{A}$ is the Alfv\'en crossing time of one cell $\Delta R$ at the inner edge of the simulation box.
This allows the system to evolve without imposing any arbitrary
conditions on $B_{\phi}$, and hence on the torques.

\begin{figure*}
  \centering
  \includegraphics[width=14.3cm,angle=90]{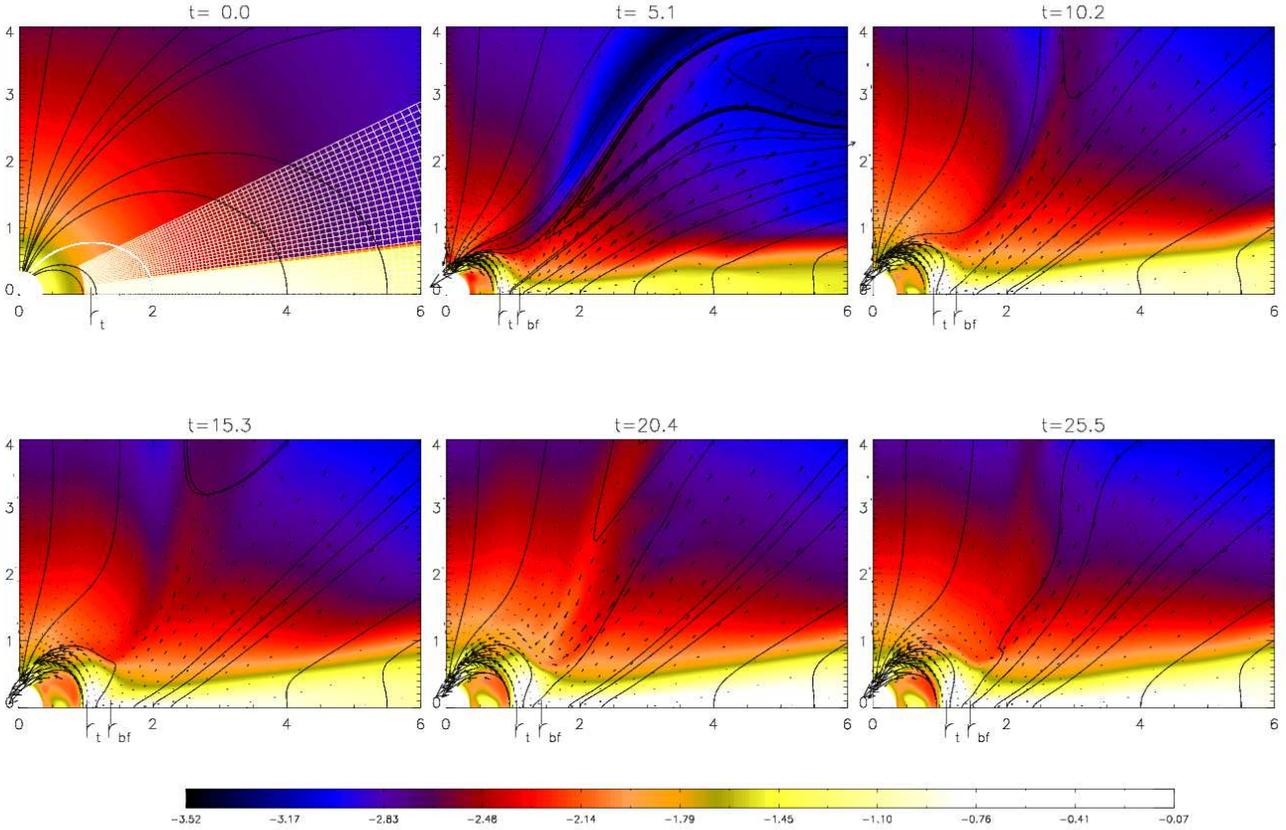}
      \caption{Resistive MHD simulation for a 5-day period CTTS 
        with $B_{*}=141\mathrm{G}$ and $\alpha_\mathrm{m}=0.1$ after t=0,5.1,10.2,15.3,20.4,25.5 Keplerian
        periods at the disc inner edge corresponding to a physical
        time of 1.5 months. We show the density distribution in the computational domain
        using a log scale. The black lines draw the
        magnetic field lines and  the black arrows represent the
        velocity field. The white line on the first snapshot represents an initial magnetic
        field line anchored at $r_\mathrm{co}$. We superimpose a
        part of the computational grid to show the good resolution 
        near the truncation radius. An accretion column is formed
        between $r_\mathrm{t}$ and $r_\mathrm{bf}$ (see text) and one
        observes the expansion of the poloidal magnetic field and 
        transient disc ejecta. The accretion rate at the stellar
        surface is equal to $1.9 \times 10^{-9}\mathrm{M_{\odot}.yr^{-1}}$
        at t=5 and stabilises around $0.91 \times 10^{-9}\mathrm{M_{\odot}.yr^{-1}}$ at t=15. 
        No X-winds are formed and the star is being spun up.} 
       \label{fig1}
   \end{figure*}

\subsection{Initial condition}

In the initial conditions we take a Keplerian disc surrounded by an adiabatic corona in hydrostatic (non rotating) 
equilibrium. The disc is adiabatic with an index $\gamma=5/3$ and an aspect ratio $\epsilon=h/r=C_s/V_K= 0.1$.
The surface of the disc is determined by the pressure equilibrium between the disc and the corona, while the initial truncation 
radius has been chosen arbitrarily.
The density and pressure expressions for both the disc and the corona are
\begin{eqnarray}
\rho_d &=& \rho_{d0} \left[\frac{2}{5\epsilon^2}\left(\frac{R_0}{\sqrt{r^2+z^2}}-\left(1-\frac{5}{2}\epsilon^2\right)\frac{R_0}{r}\right)\right]^{\frac{1}{\gamma-1}} \nonumber \\
P_d&=& \epsilon^2 V_{K0}^2 \rho_{d0}^{1-\gamma} \rho_d^\gamma \nonumber \\
\rho_c& =& \rho_{c0}\left(\frac{R_0}{\sqrt{r^2+z^2}}\right)^{\frac{1}{\gamma -1}} \nonumber \\
P_c&=& \frac{2}{5} V_{K0}^2 \rho_{c0}^{1-\gamma} \rho_c^\gamma
\end{eqnarray}
The density contrast between the disc and the corona is $\rho_{c0}/\rho_{d0} = 0.01$.
A pure dipolar magnetic field is set up in the computational domain in equipartition with the thermal pressure of the disc at $r=R_0$.
The rotation period of the star is set in order to place the corotation radius at $r=2R_0$.
The disc has no viscosity but is resistive with an alpha-like magnetic
resistivity decreasing on a disc scale height, namely 
$\nu_\mathrm{m}= \alpha_\mathrm{m}\Omega_\mathrm{k}h^{2}
exp(-(\frac{z}{h})^{4})$.  In this paper, we fix $\alpha_\mathrm{m} =0.1$.
The initial poloidal flow within the disc is zero and the disc is
slightly sub-Keplerian with $v_\phi=\sqrt{1-\frac{5}{2} \epsilon^2} \sqrt{\frac{GM_*}{r}}$.

The grid and stellar rotation period were chosen to allow good resolution at the
truncation radius $R_0$ while maintaining the co-rotation radius at $r=2R_0$ well inside the domain. 
Our polar grid of $N_R\times N_\theta=170 \times 100$ is stretched in the
spherical $R$ direction (see Fig.~\ref{fig1}) and goes from
$R_\mathrm{min}=0.35 R_0$ at the stellar surface to $R_\mathrm{max}=10.35 R_0$.

The results are presented in adimensional units: lengths are given in units of $R_0$, which corresponds to the truncation
radius of the reference simulation (see Sect. 3.5); speeds are expressed in units of the 
Keplerian speed $V_{K0} = \sqrt{GM_*/R_0}$ and densities in units of $\rho_{d0}$, which is the initial disc density at $(r=R_0, z=0)$. 
Time is given in units of the Keplerian period at $R_0$, i.e. $t_0 = 2 \pi R_0/V_{K0}$.
Mass accretion rates is given in units of  $\dot{M_0}=\rho_0 V_{K0} R_0^2$ while we express the torques in units of 
${\dot L_0}=\rho_0 V_{K0}^2 R_0^3$.

\subsection{Reference values}

We consider a $M_* = 0.8 M_{\odot}$ young star with a radius of $R_* = 2 R_{\odot}$ and a pure stellar dipole field with $B_*=141 \mathrm{G}$. 
With these assumptions the normalisation units will be $R_0=2.86R_*=4\times10^9 \mathrm{m}$, 
$V_{K0}=\sqrt{GM_*/R_0} \simeq 1.63\times10^5 \mathrm{m~s^{-1}}$ and $\rho_{d0} = \left(B_* ^2R_*^6\right)/\left(2\mu_0 \epsilon^2 V_{K0}^2 R_0^6\right) 
\simeq 5.51\times10^{-10}\mathrm{kg~m^{-3}}$. Since we place the corotation radius at $2R_0$, the rotation period of the star is $P_* = 2\pi \sqrt{8R_0^3/GM_*} \simeq
5.1$ days while the time will be scaled in units of $t_0 = 2 \pi
R_0/V_{K0} = 1.54 \times10^5 \mathrm{s} \simeq 1.78$ days. The
computational domain extends up to 0.3 AU. Finally the normalisation for the accretion rates is
given by ${\dot M_0}=\rho_0 V_{K0} R_0^2 \simeq 2.27\times10^{-8}\mathrm{M_\odot~yr^{-1}}$.

From these reference values, simulations done here can be scaled for
another range of parameters ($R_*$,$M_*$,$B_*$) in the following way:  \\
$R_0=4\times10^9 \left(\frac{R_*}{2R_{\odot}}\right)$ m, \\
$V_{K0}= 1.63\times10^5 \left(\frac{M_*}{0.8 M_{\odot}}\right)^{1/2}
                  \left(\frac{R_*}{2R_{\odot}}\right)^{-1/2} \mathrm{m~s^{-1}}$, \\
$\rho_0= 5.51\times10^{-10} \left(\frac{R_*}{2R_{\odot}}\right) \left(\frac{B_*}{141 G}\right)^2 
                       \left(\frac{M_*}{0.8 M_{\odot}}\right)^{-1}  \mathrm{kg~m^{-3}},$  \\
$P_* = 5.1 \left(\frac{R_*}{2R_\odot}\right)^{3/2} \left(\frac{M_*}{0.8M_\odot}\right)^{-1/2}$ days, \\                        
$t_0 = 1.54 \times10^5 \left(\frac{R_*}{2R_\odot}\right)^{3/2} \left(\frac{M_*}{0.8M_\odot}\right)^{-1/2}$ s, \\    
${\dot M_0}=  2.27\times 10^{-8} \left(\frac{R_*}{2R_{\odot}}\right)^{5/2}  \left(\frac{B_*}{141 G}\right)^2 \left(\frac{M_*}{0.8
  M_{\odot}}\right)^{-1/2} \mathrm{M_\odot~yr^{-1}}.$

\subsection{Methodology}

For these fixed stellar and accretion disc parameters, we make three simulations with different initial truncation radius $r_{to}$. Our reference simulation (s1) corresponds to an initial truncation radius fulfilling Eq.~(\ref{eq:rt}), namely $r_{to}=1$ with $\beta(r_{to})=1$.  Simulation (s2) is done for $r_{to}=1.5$ with $\beta(r_{to})=10$ and simulation (s3) for $r_{to}=0.5$ with $\beta(r_{to})=0.1$. 

We then let the system evolve and observe whether or not the real disc truncation
radius $r_\mathrm{t}$ converges towards the theoretical radius $r_{t,th}$ as given by Eq.~(\ref{eq:rt}). With our normalised quantities, the truncation radii in simulations (s2) and (s3) are thus expected to converge towards 1, with $\beta (r_\mathrm{t}) \sim 1$. To check this, we identify this radius and then compute the plasma beta. It is however not straightforward to define this radius since the magnetic field is not a solid wall : the radial to vertical deviation of the flow is quite smooth. In practice, we get the truncation radius $r_{t}$ by detecting a steep decrease in density at the disc midplane (see Fig. \ref{fig5}).  

\subsection{Results}

Figure~1 shows a series of snapshots of our reference simulation
(s1). The upper left panel shows the initial condition. We have
superimposed a part of the computational grid to show the resolution
achieved. We have a resolution of 8 points in the vertical direction
within the disc at each radius, while the resolution within the
accretion column reaches 20 points at the stellar surface. Magnetic
field lines are in black while the white line traces a magnetic field
line connecting the star to the co-rotation radius $r_\mathrm{co}=
2$. After a rapid transient phase with the opening of stellar field
lines, a quasi-steady situation is achieved where quasi-steady
accretion columns are formed, even with a low stellar dipole
field. The final (equilibrium) truncation radius is $r_\mathrm{t} \sim
1.1-1.2$, and thus well below the co-rotation radius.

\begin{figure}
  \centering
\includegraphics[width=7cm,angle=90]{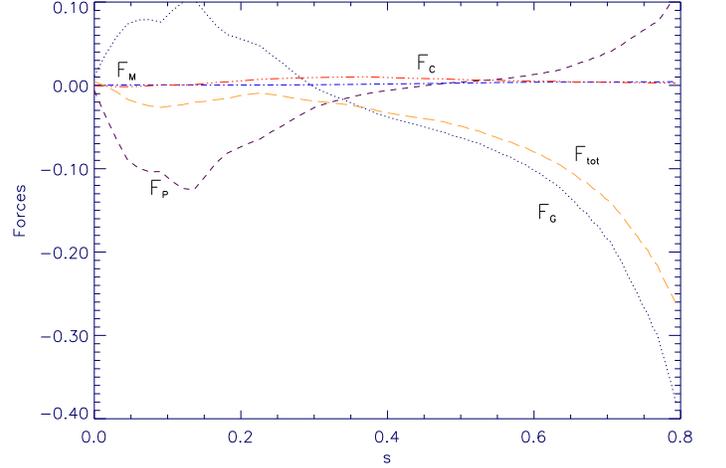}
\caption{Projection of the forces in normalised units along a magnetic field line in the middle
of the accretion column, for run (s1) at $t=10$. We represent the gravity $F_G$, the centrifugal force
$F_C$, the thermal pressure gradient $F_P$, the poloidal magnetic force $F_M$ and the
total force $F_{tot}$ as a function of the curvilinear coordinate s. The disc midplane is located at $s=0$
and the disc surface corresponds to $s \sim 0.2$. Gravity begins to dominate the dynamics only at some distance between the disc and the star ($s \sim 1.2$ at the stellar surface).}   
\label{funnel1}
\end{figure}

Figure~\ref{funnel1} shows the projection $F= \vec B_p \cdot \vec
F/|\vec B_p|$ of the various forces along a magnetic field line
located at the centre of the accretion column at $t=10$. Since the
stellar magnetic moment is directed northwards, a negative force is
actually pulling material upwards.  Accretion is achieved because of
the negative magnetic torque (not shown here) but against the poloidal
magnetic force $F_M$ which tends to prevent it. Note however that it
is
 negligible with respect to the other forces. What drives the poloidal
 motion in the accretion column is actually the plasma pressure
 gradient $F_P$ which was built in by the accumulation of accreting
 mass and allows it to be lifted up and loaded onto closed stellar field
 lines. It acts exactly as in accretion-ejection structures, enabling
 the necessary transition from the resistive MHD disc to the ideal MHD
 columns \citep{fp95}. The plasma pressure gradient remains dominant
 well above the disc surface, located around the curvilinear
 coordinate $s=0.2$.  Given the dipole topology, this is not
 surprising as material must first be lifted against gravity ($F_G$ is
 initially positive) by $F_P$. Then, at some point (which depends
 mostly on the dipole geometry) gravity overcomes it and becomes the leading agent. Again, this is only possible because the centrifugal term $F_C$ plays almost no role due to the azimuthal magnetic braking. Matter then reaches the star in a dynamical time scale with approximately free-fall velocities ($v_\mathrm{pol} \approx 310 \mathrm{km s^{-1}}$).   

The strong differential rotation beyond $r_\mathrm{t}$,  in a region where $\beta \gg 1$, leads to an expansion of the poloidal magnetic field lines. Such an expansion starts at the inner regions and enforces the outer field lines to inflate as well (but the cause there is not the differential rotation). Once these loops (inflated lines) reach the outer boundary of the computational domain, they open, mimicking a reconnection. This opening is actually an effect of the boundary conditions used but, for all practical means, we see no strong bias on the evolution of the system. Anyway, on quite short time scales ($t \sim 3$), most of the stellar magnetic flux not related to the accretion funnels around $r_\mathrm{t}$ has been opened (Fig.~1). 
One might think that a disc wind is driven in this region of the disc as the \citet{BP82} criterion is fulfilled. Moreover, some mass is indeed leaving the disc along these open field lines. But this
is only a breeze and not a proper jet: the ejected material does not
reach super-Alfv\'enic speeds. We note also that no X-winds
\citep{shu} are obtained despite the favourable magnetic configuration 
\footnote{This is a direct simulation of the scenario invoked by these authors
since the disc magnetic field comes from the opening of the magnetosphere.}. The reason why no disc wind is obtained, neither extended nor X-wind, is that the field threading the disc in these regions is far below equipartition.   

\begin{figure*}
  \centering
\includegraphics[width=9.1cm,angle=90]{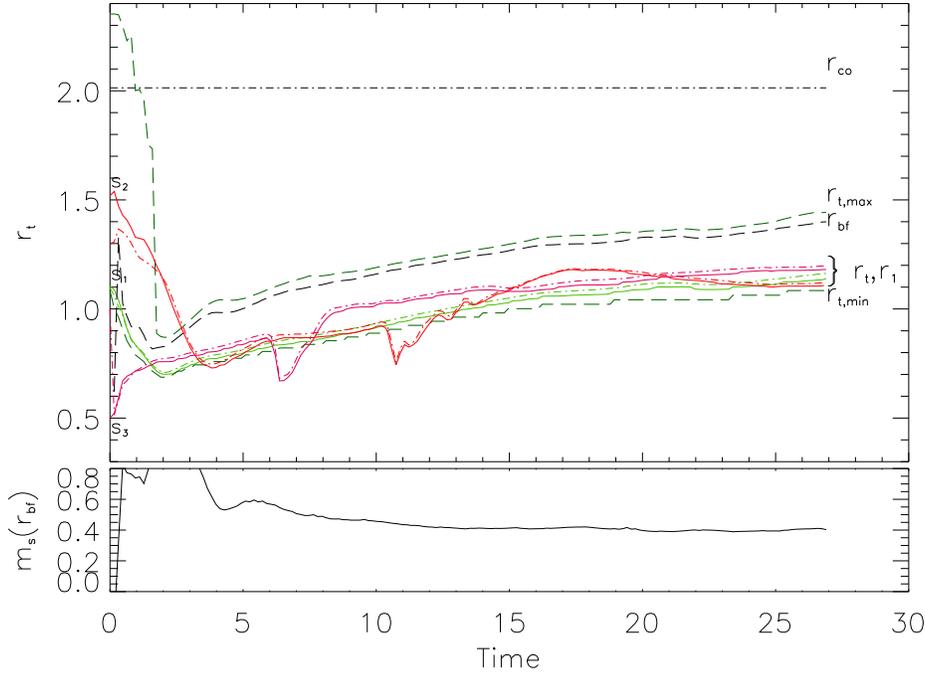}
\caption{Top : Evolution in time of the position of the truncation
  radius $r_{t}$ (solid lines) for a set of resistive MHD simulations
  with $B_{*}=141\mathrm{G}$, $P_{*}=5.1$ days and
  $\alpha_\mathrm{m}=0.1$ and a different initial truncation
  radius. The time unit is the Keplerian period at r=1. All runs
  converge towards a truncation radius $r_\mathrm{t} \sim 1.2$
  (solid lines). Dash-dot lines represent, for each simulation, the
  radius $r_1$ where  $\beta = 1$. We report also the radius
  $r_\mathrm{bf}$ (dashed line) which indicates the base of the funnel
  flow, $r_\mathrm{t,max}$ (dashed line) and $r_\mathrm{t,min}$
    for our reference run (s1).
Bottom : Evolution in time of the sonic Mach number at the base of the funnel, $m_\mathrm{s} (r_\mathrm{bf})$, for run ($s_1$).}
\label{fig2}
\end{figure*}
 
The results of varying the initial truncation radius (simulations s2 and s3) are summarised in Fig.~\ref{fig2}. In both cases, we observe a rapid convergence to $r_\mathrm{t} \approx 1.1-1.2$ on a dynamical time scale. Although some fluctuations in time can be seen in $r_\mathrm{t}$, it remains strikingly close to unity. Besides, it is really close to the radius of equipartition where $\beta = 1$, $r_1$, with a relative position lower than 4\%. This is another indication of the importance of the plasma pressure gradient in defining the truncation radius and justifies the approximation $\beta(r_t) \sim 1$.
  
The accretion rate onto the star is that which is actually
observationally determined through, e.g. veiling measurements. It is
obtained here by computing the mass flux in the accretion column,
namely $ \dot M_a= \int_S \rho \vv_p \cdot d {\bf S}=-4\pi R_*^2
\int_S \rho v_R \sin \theta d\theta$. Surprisingly, it is found to
converge towards $0.91\ 10^{-9}\mathrm{M_{\odot}.yr^{-1}}$ (see
Fig.~\ref{torques}), hence a factor 10 smaller than the mean accretion
rate in CTTS. We will come back to this issue later. If we now insert this value into Eq.~\ref{eq:rt}, we find a theoretical truncation radius $r_\mathrm{t,th} = 1.38$, using $m_s=1$. This is off by 20\%, which is not bad considering our crude approximations of such a complicated problem.

\begin{figure}
  \centering
\includegraphics[width=7cm,angle=90]{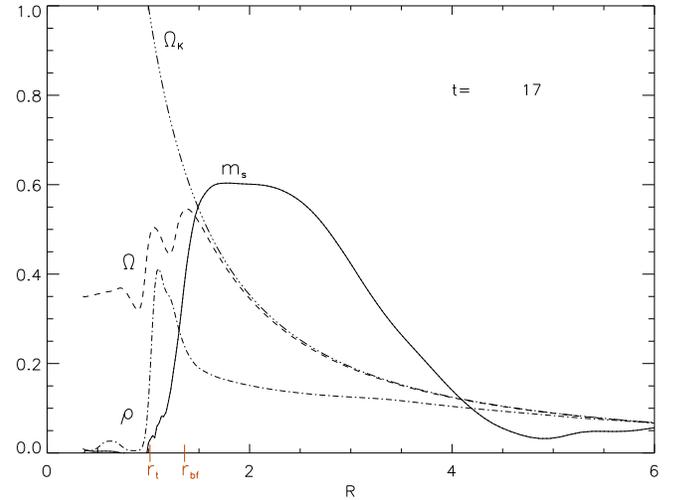}
\caption{Radial distributions of density $\rho$, angular velocities $\Omega$ (real) and $\Omega_K$ (Keplerian) and sonic Mach number $m_s= u_r/C_s$ at the disc midplane, for run (s1) after 17 Keplerian rotations. Notice the large accretion velocity in the region of opened stellar field lines.}
\label{fig5}
\end{figure}

\begin{figure}
  \centering
\includegraphics[width=7cm,angle=90]{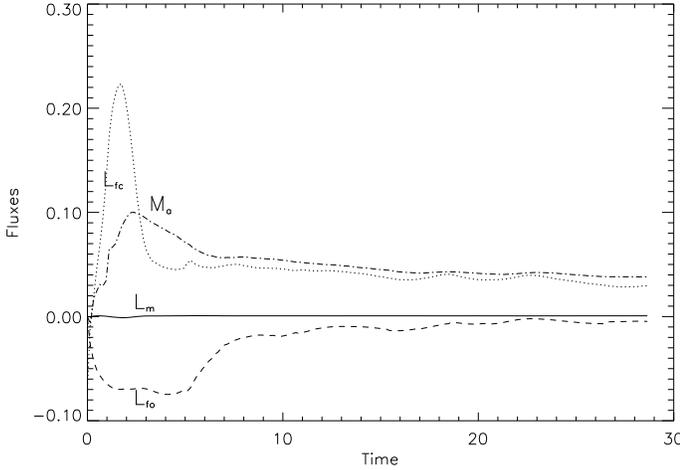}
\caption{Evolution in time of the accretion rate $\dot M_a$ (dash
  dotted line), angular momentum flux transported by matter $\dot L_m$ (solid line)
and by the magnetic field, computed for the closed  $\dot L_{fc}$ (dash line)
and open $\dot L_{fo}$ (dotted lines) field lines. The units are normalised and the two hemispheres are taken into account.}
\label{torques}
\end{figure}

Since  $\beta(r_t) \sim 1$ is well verified, the main source of discrepancy in Eq.~\ref{eq:rt} is due to the assumption of $m_s(r_t) \sim 1$. This is too crude for an obvious reason. Indeed, the disc truncation radius, as measured by the steep drop in density at the equatorial plane, is actually the point where $u_r=0$, hence $m_s=0$. Figure~\ref{fig2} shows the evolution of the sonic Mach number $m_s$ computed for run (s1) at $r_\mathrm{bf}$, which corresponds to the base of the funnel flow where the poloidal magnetic pressure matches the poloidal ram pressure. It can be seen that $r_{bf}$ is larger than $r_t$ by almost 30\%. This is not surprising as it is necessary to first brake efficiently material ($r_\mathrm{bf}$) before being able to lift it up ($r_t$). This is also illustrated in Fig.~\ref{fig5} where we plot the radial profile of several quantities at the disc midplane: density $\rho$, angular velocities (real $\Omega$ and Keplerian $\Omega_K$) and $m_s$. Clearly, taking $m_s =1$ to derive $r_{t,th}$ is too crude. We find that using a value of $m_s \simeq 0.45$, namely close to the real value $m_s(r_{bf})$ (see Fig.~\ref{fig2}), provides a much better estimate with $r_{r,th}$ close to $r_t$ with an accuracy better than 10\%. 

Although the simulations here do not take into account viscosity, for
completeness, we plotted in Fig.~\ref{fig2} the evolution of the radius $r_\mathrm{t,max}$ for run (s1). To do so, we assumed $\alpha_\mathrm{v} =1$ (thus overestimating the viscous torque) and measured numerically $q$ (see Sect.~2). As expected, $r_\mathrm{t,max}$ remains always significantly larger than the real truncation radius $r_\mathrm{t}$, which implies  that criterion A is not good enough. On the other hand, criterion B would give a truncation radius located at $r_t = 0.3$: no accretion column should have been observed at all in our simulations. Our results clearly show that criterion B is not relevant. What about the criterion as expressed in Equation~(\ref{eq:rtmin})? We plotted in Fig.~\ref{fig2} the evolution of the radius $r_\mathrm{t,min}$ given by this equation and computed for run (s1). It turns out that it gives a good (though under-) estimate of the real truncation radius $r_t$ as already pointed out by \cite{rom}. This is because the star disc interaction introduces a sharp decrease in both the disc midplane density $\rho$ and azimuthal velocity $v_\phi = \Omega r$ (see Fig.~\ref{fig5}). However, as stressed in Sect.~2, while basically correct, such a criterion is meaningless as a predictive tool.

As far as the formation of funnel flows is concerned, we fully confirm the results of \citet{rom}: these flows are indeed robust features of axisymmetric MHD simulations. Their different boundary conditions on the magnetic field at the stellar surface, namely a fixed normal component and free conditions for $B_\theta$ and $B_\phi$, do not finally play any significant role for the truncation of the disc and the formation of accretion columns. We have also done simulations with other values of the magnetic resistivity parameter $\alpha_m$ with no change in the truncation radius. Note also that \citet{rom} have included viscosity in their simulations while we did not, with no significant difference in the location of the truncation radius. To be more specific, using Eq.~(\ref{eq:rt}) with the higher accretion rate of $\dot M_a = 9 \times 10^{-8}
M_\odot.yr^{-1}$ as measured by \citet{rom}, we find $r_\mathrm{t,th} =
0.9$ for $B_{*}=1.1\mathrm{kG}$ using $m_s =0.45$, which is consistent with the truncation radius shown Fig.~16 in
\cite{rom} with a
good accuracy.  Furthermore, using Eq.~(\ref{eq:rt}) with the values
provided by \citet{kuker} one gets truncation radii smaller than the
inner radial boundary, which explains why these authors did not find
accretion columns. We are therefore confident on our main conclusion,
that is the validity of our criterion (\ref{eq:rt}).

\begin{figure}
  \centering
\includegraphics[width=7cm,angle=90]{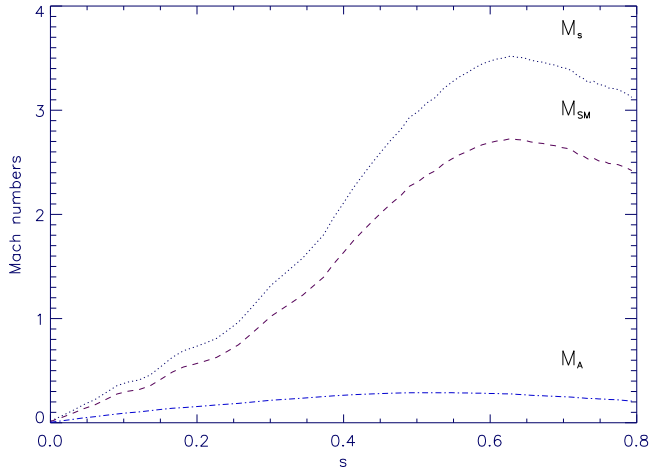}
\caption{Sonic (dotted line), slow-magnetosonic (dashed line) and
  Alfv\'enic (dash-dotted line) Mach numbers within the funnel flow along the same
  magnetic field line at t=10 as in Fig. \ref{funnel1}. }
\label{mach}
\end{figure}

We now consider the fluxes of angular momentum, namely ${\dot L_m}$ carried in by the infalling material and  ${\dot L_f}$ by the magnetic field
 \begin{eqnarray}
\dot L_m &=& \int_S \rho \Omega r^2 \vv_p \cdot d {\bf S}= -4\pi R_*^2
\int_S \rho \Omega r^2 v_R \sin\theta d\theta \nonumber \\
\dot L_f &=& - \int_S r B_\phi \BB_p \cdot d{\bf S}= 4\pi R_*^2
\int_S r B_\phi B_R \sin\theta d\theta 
\end{eqnarray}
To explicitly write these expressions we assumed that the surface element is directed inwards with respect to the surface of the star.
The flux carried by the magnetic field is the sum of that carried by
closed field lines and open field lines, namely  $\dot L_f= \dot
L_{fc} + \dot L_{fo}$. It actually corresponds to two possible
electric circuits each related to two different electromotive
forces. A positive flux describes a positive torque acting on the star
and leading to a spin up while a negative flux leads to a spin
down. The time evolution of these fluxes is shown in
Fig.~\ref{torques} for our reference run (s1). Not surprisingly, the
incoming angular momentum flux $\dot L_m$ due to the accreting
material is positive but totally negligible with respect to that
carried by the closed magnetic field lines $\dot L_{fc}$, which is positive as well: the accreting star is only being spun up. 
When looking more closely at $\dot L_{f}$, it turns out that there is a
negative magnetic contribution to the torque due to the open field
lines $\dot L_{fo}$. This has also been previously reported in
simulations and is a natural outcome of the star-disc interaction
\citep{Long05}. We stress however that the actual torque is not
controlled and one should not take it at face value. Indeed, this
region of the magnetosphere should be the locus of a stellar wind but
the physics of its launching has not yet been addressed.  
Finally, let us turn back to our result that the accretion rate $\dot
M_a$ measured onto the star is about 10 times smaller than
expected. In fact, $B_*$ and $\dot M_a$ were considered  in
Eq.~\ref{eq:rt} as independent parameters. However, the mass inflow at
the inner edge of the accretion disc is constrained by the magnetic
topology. Basically, the magnetosphere acts as a nozzle and the mass
flow cannot be arbitrary. In a strict steady-state analysis it would
be imposed by the regularity condition at the slow magnetosonic
point. In our simulations, the flow reaches the slow point $M_{SM}=1$ above the disc at $s \sim
0.3$ (see Fig. \ref{mach}). Then, it  
remains supersonic with a maximum for
the sonic Mach number around $M_s \sim 3.5$ but the funnel flow is
always sub-Alfv\'enic with a maximum Alfv\'enic Mach number $M_A \sim
0.3$ near the middle path between the stellar surface and the disc inner edge.

\section{Conclusion}

We have confirmed that the formation of accretion funnel flows is a robust feature of axisymmetric star-disc interactions. We investigated the physical conditions required to produce steady-state funnel flows onto a dipole and provided an analytic expression of the disc truncation radius. We then used MHD simulations with VAC to show its validity.

Our theoretical expression $r_{t,th}$ relates the disc truncation
radius to astrophysical parameters such as stellar dipole field
$B_*$, mass $M_*$, radius $R_*$ and accretion rate $\dot
M_a$. Although it resembles the Alfv\'en radius \citep{Elsner77} sometimes invoked in similar situations \citep{Bou07}, it has been physically motivated on very different ground. It is shown that it gives an accurate prediction of the real truncation radius as obtained in current 2D MHD simulations of a star-disc interaction.

We report MHD simulations displaying accretion funnels with a weak stellar dipole
field $B_* \sim 140$ G from a resistive non-viscous disc. In this case, the disc inner edge is found closer to the star, below the co-rotation radius, in agreement with the size of inner disc holes. However, the accretion rate onto the star which is imposed by the physics of magnetic accretion is measured to be
 ${\dot M} \sim 10^{-9} M_\odot.yr^{-1}$. Even though this magnitude
 of accretion rate is found for some T~Tauri stars
 \citep[see][]{Gullbring98}, this result shows that it is necessary to
 have stronger fields (by a factor of around 7) to have magnetospheric
 accretion for typical mean accretion rates of  ${\dot M} \sim 10^{-8}
 M_\odot.yr^{-1}$. On the other hand, the few circular polarization
 measurements available, provide an upper limit of 100-200 G for the
 dipolar component. 
However, recent spectro-polarimetric observations coupled with magnetic field
reconstruction conducted on 2 CTTS  V2129 Oph and BP Tau provide
higher dipole field components of 350 G and 1.2 kG respectively
\citep[Donati et al. 2007b in prep.]{Donati07a}. In any case, if the presence of kG dipole
fields around CTTS are indeed ruled out, the requirement of forming quasi-steady accretion funnels imposes that the magnetic topology must be different. This is a firm result based on our dynamical calculations. 
 One then needs to consider other stellar magnetic field topologies such as multipolar components and/or an inclined dipole configuration \citep{rom03,Long07}. 
In this paper, since there is no viscosity on our resistive accretion
disc, we privileged a situation where  $r_\mathrm{t} \ll
r_\mathrm{co}$ to reach an open magnetic topology within the main part
of the disc. This made it possible to have a braking torque feeding the disc even without viscosity. Apart from the localised magnetic flux channeling the accretion funnel, the other stellar magnetic field lines become causally disconnected from the disc. 
This forbids any disc-locking mechanism and the star is being spun up
as a consequence of accretion. We also report the non development of
X-winds despite the favourable topology. Both of these aspects deserve
however a more detailed analysis as they may depend on the disc
resistivity. While the disc truncation radius remained close to its
predicted position for about 25 Keplerian periods at the disc inner
edge, some variability is obviously taking place. This is a very
promising topic as veiling measurements probing accretion onto the
star do show variability on different time scales
\citep{Alencar00}. However making longer simulations requires one to implement
the turbulent viscosity to allow for accretion within the disc beyond
corotation. This work is in progress.

\acknowledgements{
We acknowledge useful suggestions and constructive criticism of an anonymous referee on a first version of this paper.
We thank the computational facilities of the Service  Commun de Calcul Intensif de l'Observatoire de Grenoble (SCCI). 
The present work was supported in part by the European CommunityÕs Marie Curie Actions - Human Resource and Mobility within the JETSET (Jet Simulations, Experiments and Theory) network under contract MRTN-CT-2004 005592.
}

\end{document}